\nonstopmode
\documentclass[ras]{agu2001}
\usepackage{graphicx}

\authorrunninghead{KESTEVEN ET AL.}
\authoraddr{M.KESTEVEN,
AUSTRALIA TELSECOPE NATIONAL FACILITY, CSIRO, AUSTRALIA,
(michael.kesteven@csiro.au)}

\titlerunninghead{ADAPTIVE FILTERS FOR PULSAR OBSERVATION}

\begin{document}

\title{Adaptive Filters Revisited - RFI Mitigation in Pulsar Observations}

\authors{M. Kesteven, \altaffilmark{1}
G. Hobbs, \altaffilmark{1}
R. Clement,\altaffilmark{1,2}
B. Dawson,\altaffilmark{1}
R. Manchester,\altaffilmark{1}
T. Uppal \altaffilmark{1} }

\altaffiltext{1} {Australia Telescope National Facility, CSIRO}
\altaffiltext{2} 
{Dept. of IT and Electrical Engineering, University of Queensland}

\begin{abstract}

Pulsar detection and timing experiments are applications where
adaptive filters seem eminently suitable tools for
radio-frequency-interference (RFI) mitigation.  We describe a novel
variant which works well in field trials of pulsar observations
centred on an observing frequency of 675\,MHz, a bandwidth of 64\,MHz
and with 2-bit sampling.
Adaptive filters have generally received bad press for RFI mitigation
in radio astronomical observations with their most serious drawback
being a spectral echo of the RFI embedded in the filtered signals.
Pulsar observations are intrinsically less sensitive to this as they
operate in the (pulsar period) time domain.  The field trials have
allowed us to identify those issues which limit the effectiveness of
the adaptive filter.  We conclude that adaptive filters can
significantly improve pulsar observations in the presence of RFI.

\end{abstract}

\begin{article}

\section {Introduction}

The days of interference-free observations in radio astronomy are now
long gone.  Increasingly, observations such as the search for
red-shifted HI will need to be made outside the bands allocated to
radio astronomy.  There are also substantial pressures from
commercial, defence and other interests for greater access to the
radio-frequency spectrum.  This means that the radio astronomers can
no longer rely on the regulatory authorities for an interference-free
environment; we need to explore the possibilities for co-existence.

\begin{figure*}
\includegraphics[width=20pc, angle=-90]{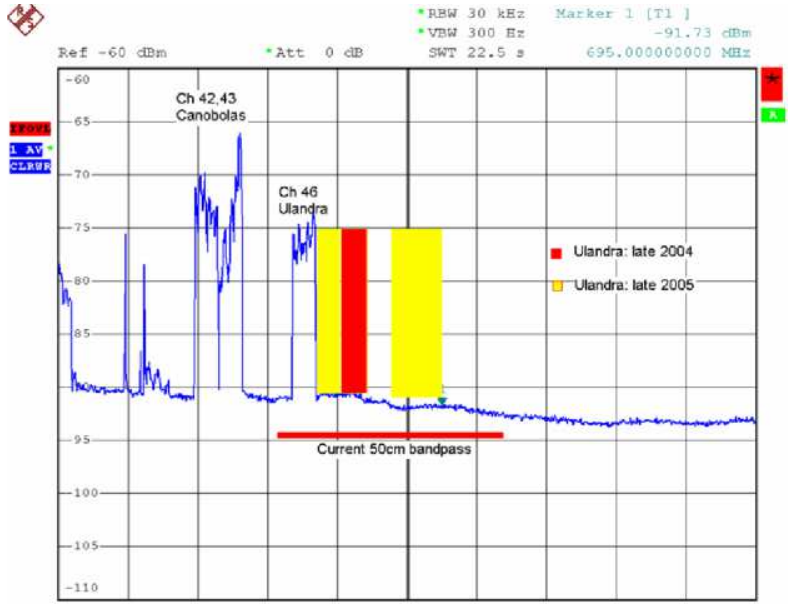}
\caption{The RFI environment at Parkes in the 50cm band. The frequency
span in the image is 200 MHz.}
\label{pksrfi}
\end{figure*}

The work described in this paper was prompted by the commissioning of
a new digital TV transmitter located on Mt. Ulandra, about 200 km south of
the Parkes observatory.  This transmitter operates at a frequency of
645 MHz which is within the bandpass of one of the receivers used for
pulsar observations.  Figure~\ref{pksrfi} shows the current RFI
environment at Parkes at these frequencies and the location of three
further transmitters scheduled for commissioning in the near future.
While repositioning the receiver bandpass might avoid this particular
source of radio-frequency-interference (RFI), it is clear that the
observatory needs to continue to develop its expertise in RFI
mitigation.

The adaptive filter is one of the promising areas of interference
mitigation: the filter detects the presence of interference in the
astronomery data, and derives a suitable correction function to
remove (or at least reduce) the interference.  Several groups at the
Australia Telescope National Facility have been engaged in RFI
mitigation experiments for a number of years, concentrating in
particular on the post-correlation class of adaptive filters (Briggs
et al. 2000; Kesteven 2004).  This work has been successful in
demonstrating useful RFI mitigation in spectroscopy experiments.
While the post-correlation adaptive filter may be applicable to pulsar
observations, it will be expensive in computational resources.

In this paper, we show that the original form of the adaptive filter
is an effective and cost-efficient solution to the requirements of
pulsar observations.

\section{The Adaptive Filter}

\begin{figure*}
\noindent\includegraphics[width=30pc]{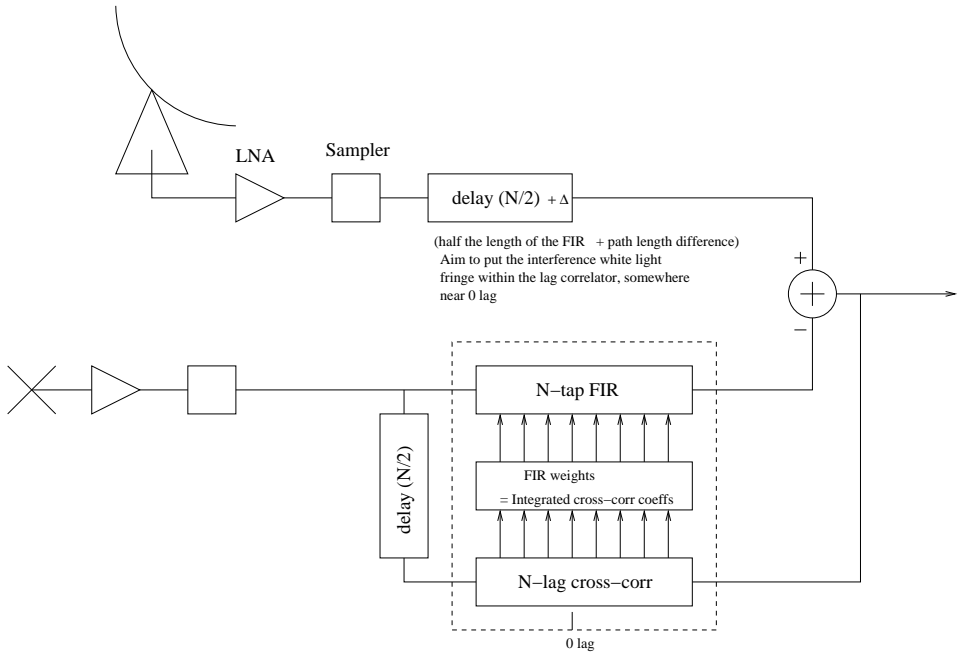}
\caption{The pre-detection filter.  We require a reference antenna which 
captures a copy of the interference while remaining insensitive to the
astronomical signal.  The filter modifies the reference antenna's signal
into a close copy of the interference in the astronomy signal.}
\label{model}
\end{figure*}

Figure~\ref{model} summarises the RFI problem to be solved and the
nature of the solution.  The astronomical antenna collects signals
from some target region on the sky.  The receiver responds to the 
astronomical signals primarily from the antenna's main beam.
The antenna also receives
interference through one of the many antenna sidelobes.  Astronomers
will find their data corrupted by this interference -- at times to
such an extent that the data are useless.

An adaptive filter is a device that can remove much of the
interference from the astronomy signal without affecting
the astronomical information.  The hardware consists of a
reference antenna, organised to be very responsive to the
interference, and to have little or no response to the astronomy
signal.  The heart of the device is a filter which acts on the
reference antenna signal to modify it into a close copy of the
interference in the astronomy channel. A subtraction will then yield a
cleaned astronomy signal that is free of interference.  The third
component of the adaptive filter is a mechanism to control the filter
in order to meet some optimising criterion. Such a device, based on a
convolutional filter, was described in detail by Barnbaum \& Bradley
(1998). It is an elegant scheme which operates directly on the 
intermediate frequency band (IF) that
the astronomer would direct to final processing stage -- the pulsar
de-dispersing, folding and timing computer in our case.

Assume, for the moment, that the system is operating in a narrow radio
frequency band.  In this case, we would require the filter to adjust
the gain and phase of the reference IF until the interference is a
good match to the interference in the astronomical channel.  A
subtraction will yield an interference-free IF.  Unfortunately the
reference IF also contains noise from its receiver.  Increasing the
gain in order to balance the RFI will allow an increasing amount of
receiver noise into the output IF.  Decreasing the gain degrades the
interference cancellation.  Since the astronomer generally does not
distinguish noise power from interference power in his spectrum, the
optimum filter gain (from the astronomer's perspective) is the setting
with the minimum additional power in the output IF.

More formally, let:

\[
P_{out} = N_{ast} + P_{rfi}\times(c_{ast} - g*c_{ref})^{2} + g^{2}N_{ref}
\]

where $N$ is the receiver noise power in the astronomy ($ast$) or
reference ($ref$) channels, $P_{rfi}$ is the interference power at the
site, $c$ describes the complex voltage coupling of the interference
into the two channels and $g$ is the complex voltage gain of the filter.

Minimising $P_{out}$ with respect to g leads to :

\[
g = \frac{c^{*}_{ast}c_{ref}P_{rfi}}{N_{ref} + c^{2}_{ref}P_{rfi}}
\]

The power at this optimal gain setting is:

\[
P_{out} = N_{ast} + \frac{c^{2}_{ast}P_{rfi}}{(1 + {\rm INR})}
\]

where INR is the ratio of the interference power in the
reference IF to the noise power,

\[
{\rm INR} = \frac{c^{2}_{ref}P_{rfi}}{N_{ref}}
\]

In other words, the astronomer will see the ``interference'' in his
spectrum reduced from P (= $c^{2}_{ref}P_{rfi}$) to P/(1~+~INR).  
With no filtering the spectrum is corrupted by interference;  with the
filter in operation the corruption is due to a small residue of
interference along with a small fraction of noise from the receiver
in the reference channel. 

The adaptive filter in practice operates over a wide frequency
range. The scheme outlined above is readily modified to suit that need
by recognising that the gain is frequency dependent and implementing
the filter as a convolutional filter.  The cross-spectrum output from
the correlator is then the Fourier transform of the correction to the
filter weights. The filter weights are optimised when the
cross-spectrum is zero.

The correlator output will of course be subject to noise fluctuations
which can be smoothed by averaging.  The appropriate averaging time is
set by the time scale on which the noise-free filter settings would
change, i.e. by the time scale on which the coupling terms change.
This is set by propagation considerations such as the relative delay
between the reference and astronomy antennas, or the changing
proportions of the multi-path propagation.  As these are relatively slow
effects, we used in the pulsar experiments a time scale of 3\,ms.

The FIR weights are computed as follows.  Let $\{w^{n}_{i}\}$ be the
set of weights in use during the n-th integration, R(t) is the voltage
in the reference channel, A(t) the voltage in the astronomical channel
and F(t) is the voltage output from the FIR filter,

\[
F_{j} = \sum_{i} R_{(j-i)} w^{n}_{i}
\]

The cross-correlation terms can be written as

\[
C_{i} = \frac{1}{N} \sum_{k} (A_{k} - F_{k}) R_{(k-i)}
\]

and the updated weights are

\[
w^{n+1}_{i} = w^{n}_{i} + \epsilon C_{i}
\]

where $\epsilon$ controls the convergence of the filter to the optimum 
setting. The loop is critically damped when $\epsilon = 1/Tsys_{ref}$.

This filter has a number of desirable qualities:

\begin{enumerate}
\item  It adapts automatically to changes in the coupling coefficients.
This is important as different sidelobes could be involved as the
antenna follows a source. The relative delay between the reference and
astronomical antennas may change if the interference source is moving (e.g., 
a satellite). It will accomodate to changes in the receiver gains.

\item The filter is robust to multi-path propagation.

\item The filter action ceases when the interference ceases.  There
is no noise penalty at low to zero interference.

\item  It can handle multiple independent sources of interference provided 
that there is no overlap in frequency.

\item  Filters can be cascaded which would allow the system to accommodate
sources which overlap in frequency. This strategy is also relevant if
the interference is so spread out in direction that multiple reference
antennas are required.

\end{enumerate}

To the astronomer's eye, the filter reduces the interference-related
noise power by an attenuation factor equal to 1/(1+INR).  The filter
starts to become ineffective when INR $\sim$ 1.

\section {The Field Trials}

The Parkes observatory has an on-line pulsar processor (known as
CPSR2) that was constructed by the Caltech and Swinburne University
groups (Bailes, 2003).  This unit streams two baseband IFs (two
polarisations, each IF sampled at 128\,MSamples/sec), each 64\,MHz
wide to a disk farm, for real-time processing by a bank of 32\,PCs.
Each PC is assigned a 1-GByte file for dedispersing, folding at the
pulsar period and timing.  Our long-term aim is to build a hardware
adaptive filter and install it ahead of the processor and other data
acquistion systems.

\begin{figure*}
\noindent\includegraphics[width=30pc]{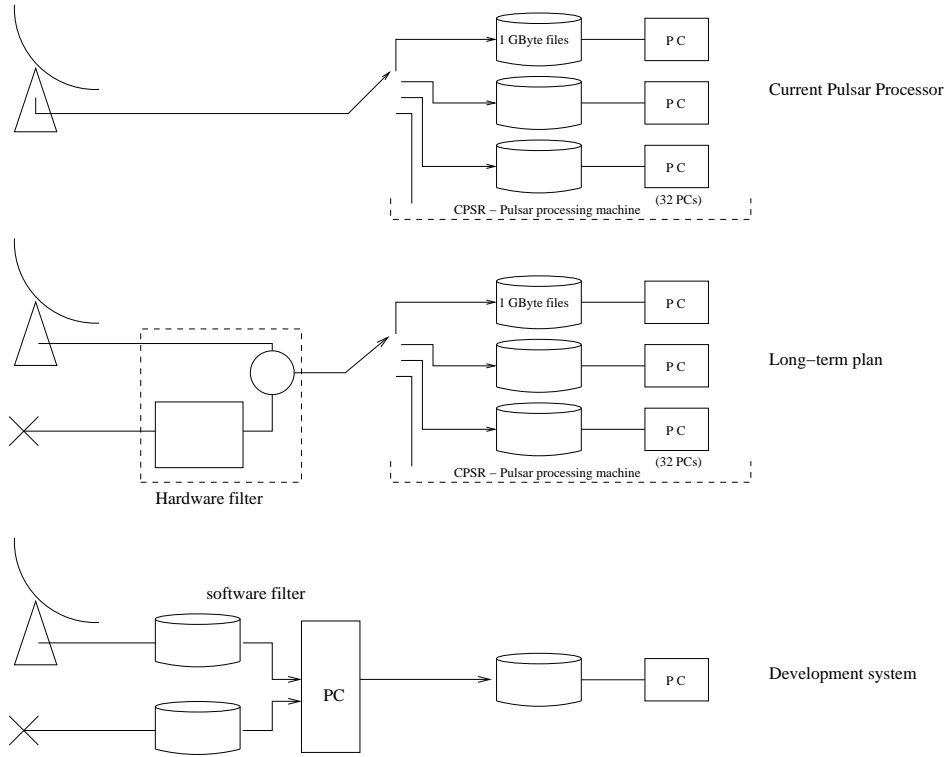}
\caption{The pulsar processing models.  The top figure shows the
current implementation.  The middle figure shows our long term aim: 
a real-time hardware filter which cleans the IF ahead of the
CPSR2 processor. The lower figure shows our development mode: a
software filter operates on a file of data, replacing it with a filtered
version for subsequent processing.  In this mode the operation is not
real time.}
\label{cpsr}
\end{figure*}

For the current experiments we exploit the CPSR2 architecture to
develop the interference rejection algorithm and explore the practical
difficulties in RFI mitigation with the use of a software filter.  The
schemes are shown in figure~\ref{cpsr}.  The software filter cannot
support real-time operation, but in every other respect it provides a
comprehensive trial.  We take two disk files, one with
dual-polarisation pulsar data, the second with the IF from the
reference antenna and we adaptively filter the pulsar data to create a
fresh file with filtered data.  This file is inserted back into the
system for standard processing.

The reference antenna is mounted on the tallest tower
at the observatory and has been oriented to maximise the signal
received from Mt. Ulandra.  The original experiments used a
yagi antenna;  this has since been replaced by a 3.5m diameter antenna.
The astronomical targets are a variety of
pulsars that are detectable in a 16-sec observation.  Most of the
results shown below are based on the millisecond pulsar
PSR~J0437$-$4715 which has a period of 5.7\,ms. The observing band
centre is set to 675\,MHz, the bandwidth to 64\,MHz and we use 2-bit
sampling.

The observations can address the following questions:

\begin{enumerate}
\item Does our filter implementation behave as predicted?
\item Does our filter introduce pulsar specific side-effects?
\item Are there other factors which limit the effectiveness of the filter
in mitigating the RFI?
\item Are there specific sampling issues that need to be addressed?
\end{enumerate}

\subsection{Filter Performance}

We have implemented in software the filter shown in
figure~\ref{model}.  Although the data is 2-bit sampled, all the
operations within the filter are 32-bit floating.  The output is
normalised and resampled to 8-bits to provide data in a format
suitable for the down-stream processing.

\begin{figure*}
\noindent\includegraphics[width=30pc]{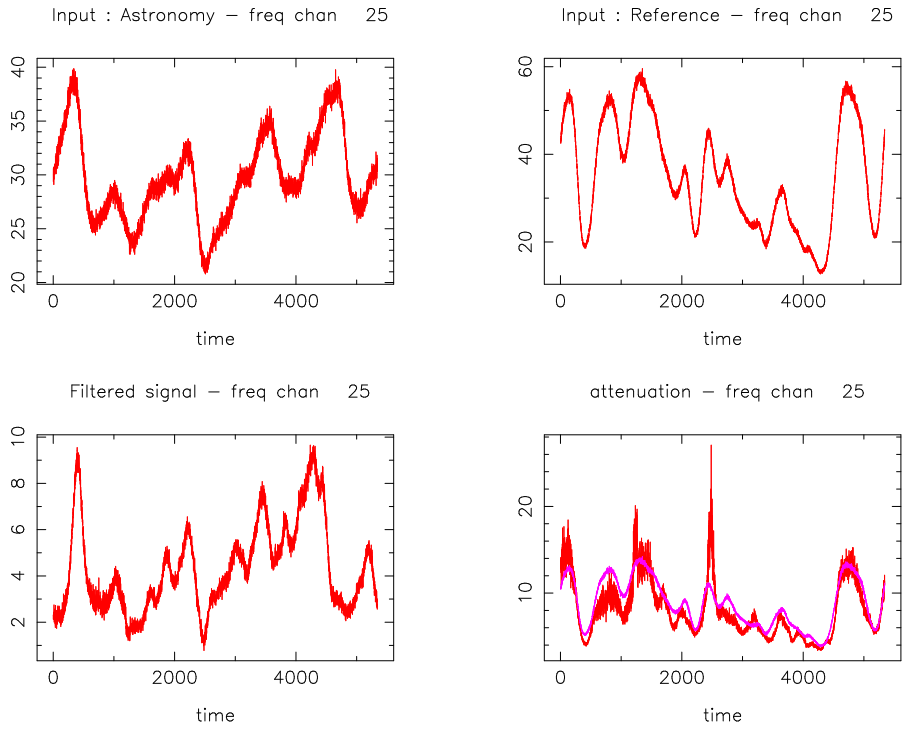}
\caption{The filter in action.  Each plot shows the filter performance in
one selected frequency channel (number 25), computed every 3 ms.  The
plot at the lower right shows the attenuation (filtered excess noise
power divided by the excess noise in the raw astronomy IF, where
``excess noise power'' is the amplitude above the RFI-free level).
The attenuation computed from the INR of the reference IF is also
shown - it is the trace which is essentially free of noise.}
\label{filt2}
\end{figure*}

The FIR filter has 128 taps (the number of delay steps in the filter),
as does the cross-correlator. The spectral resolution is a sin(x)/x
window of width 1 MHz, set by the sampling rate (128 MS/s) and the
number of taps.  The FIR weights are revised every 3\,ms. The weights
will need to be updated if there are changes in the way the
interference appears in the reference and astronomy IFs. For example,
while the main antenna is following a sidereal target its sidelobes
will track over the source of interference.  The interference may also
suffer multipath propagation to the reference antenna
leading to episodes of fading.  Our experiments suggest that at our
site the characteristic timescale for these effects (see
figure~\ref{filt2}) is measured in seconds.

\begin{figure*}
\noindent\includegraphics[width=25pc]{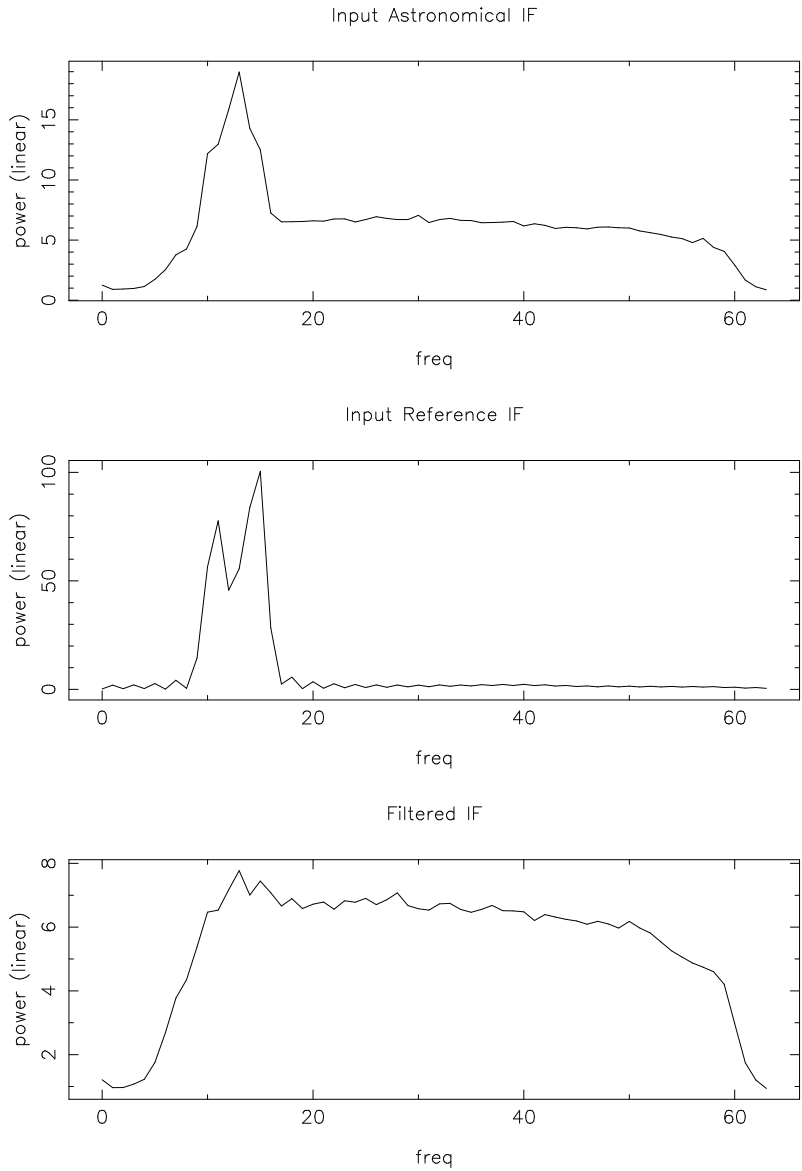}
\caption{The filter in action.  The top panel shows the raw astronomy IF,
the middle panel shows the reference IF (INR $\sim$ 50:1) and the
lower panel shows the filtered IF.}  \label{filt1}
\end{figure*}

Figure~\ref{filt1} provides an overall picture of the filter's
performance.  It shows the IF power spectra before and after
filtering.  The INR is large and the residual noise (the difference
between the filtered spectrum and the RFI-free receiver bandpass) is
low, consistent with the large INR in the reference channel.

Figure~\ref{filt2} presents a different perspective. The filter is
analysed at 3\,ms integrations during the processing.  This example
was selected to highlight a problem: during the 16\,second
observation the RFI amplitude changed in both the reference and
astronomy channels independently. We believe that this is
due to multipath propagation acting independently on the astronomy
and reference channels.  The 3\,ms update interval is
adequate to follow these changes, and the attenuation tracks the INR
as predicted.

Our observations all indicate that the filter behaves as predicted,
with a clear message that maintaining a large INR is vital.

\subsection{Are there Side-Effects?}

Our brief is to ensure that the filter does not affect the spectral,
polarisation or timing characteristics of the pulsar. Our approach has
been to do detailed comparisons of the processed data -- with and
without filtering.

\begin{figure*}
\noindent\includegraphics[width=30pc]{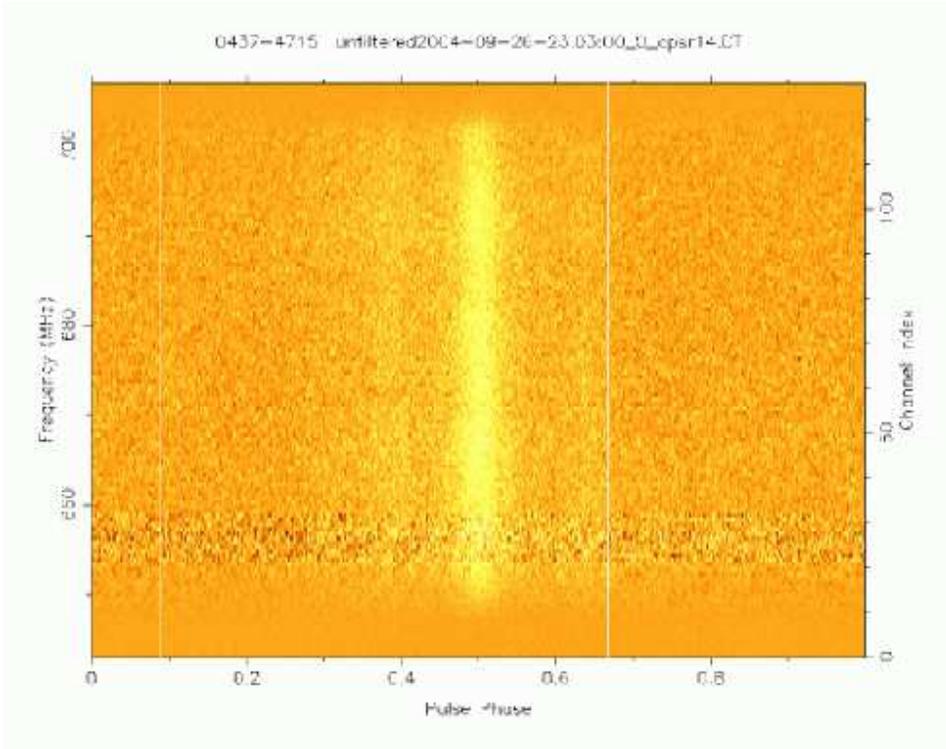}
\caption{The output from the pulsar processor after de-dispersion and folding 
at the pulsar period of a 16.7-sec segment of unfiltered data.}
 \label{filt3a}
\end{figure*}

\begin{figure*}
\noindent\includegraphics[width=30pc]{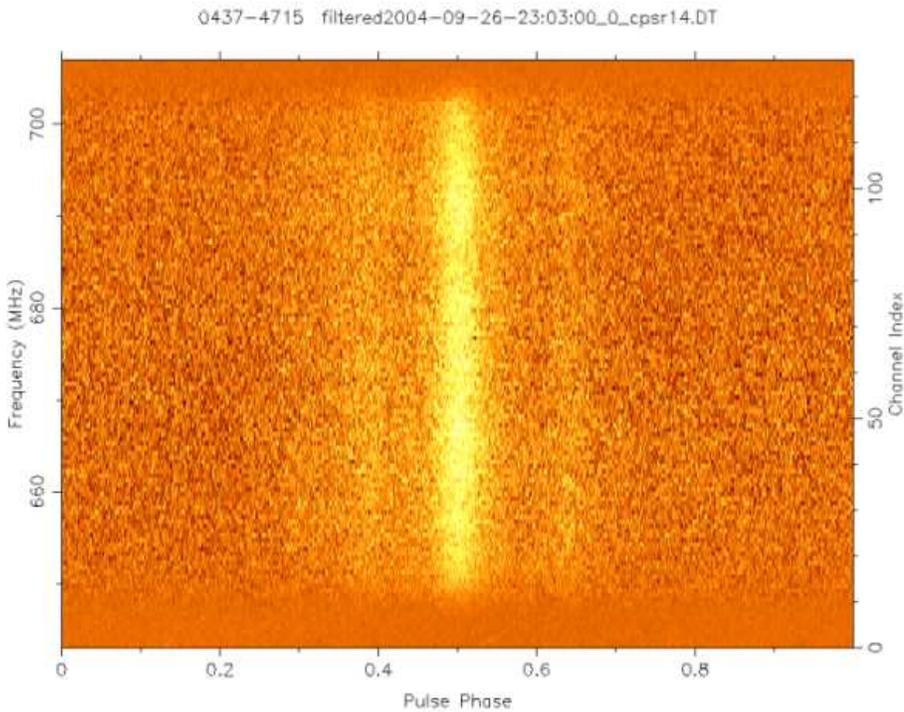}
\caption{The output from the pulsar processor after de-dispersion and folding 
at the pulsar period of filtered data.}
 \label{filt3b}
\end{figure*}

Figures~\ref{filt3a} and \ref{filt3b} show the typical raw material,
the output from the pulsar processor with data folded at the pulsar
period.  There are 2048 channels along the pulsar phase axis and 128
along the frequency axis.  The data has been de-dispersed, so the
pulsar's ``pulse'' is visible as the vertical trace near pulsar phase
0.5.  The basic question is whether the pulse in the RFI spectral
channels has the same characteristics as the pulse in the RFI-free
channels.

\begin{figure*}
\noindent\includegraphics[width=30pc]{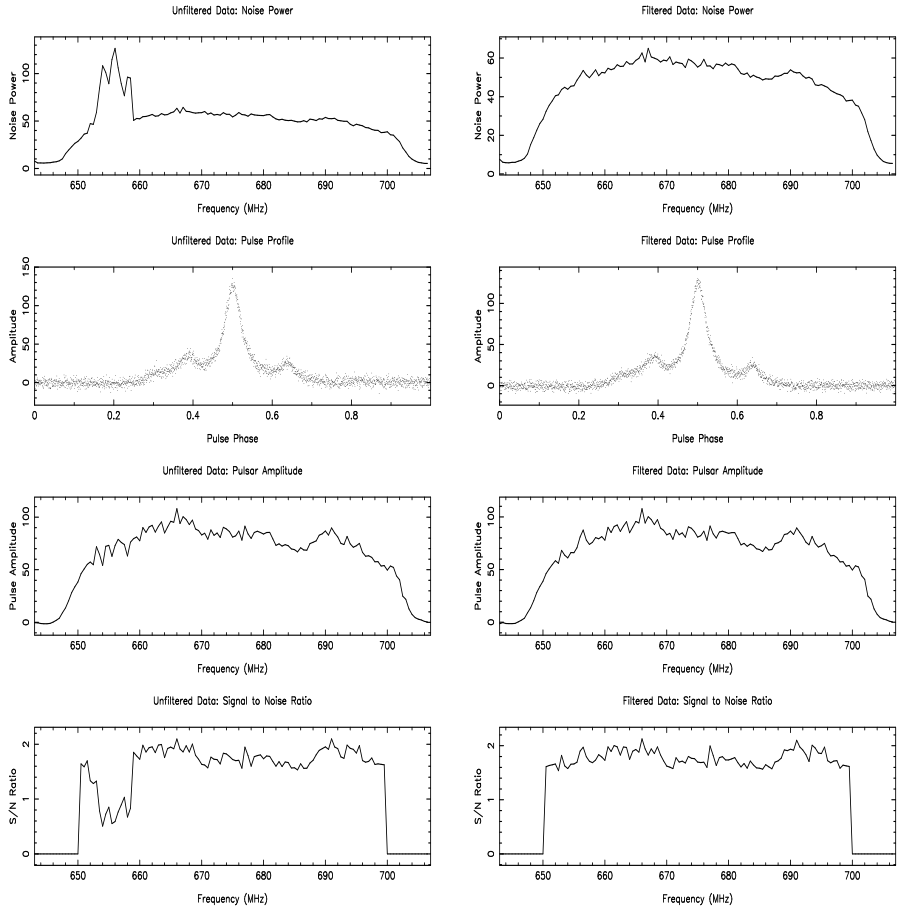}
\caption{A statistical summary of figures \ref{filt3a} and \ref{filt3b}.  
The top panel shows the mean power in each frequency channel. 
The second panel shows the pulse profile.  The third panel
shows the amplitude of the pulsar pulse, at each frequency;  it is the
power summed over a narrow window in pulsar phase centred on the pulse peak.
The fourth panel shows the Signal-to-Noise ratio of the pulse amplitude.}
\label{filt4}
\end{figure*}

Figure~\ref{filt4} provides an analysis of the data in figures
\ref{filt3a} and \ref{filt3b}.  These figures show the detected power,
after folding and de-dispersing, as a function of pulsar phase and
observing frequency ($P(\phi,f)$).  Each column in figure~\ref{filt4}
has four plots.  The first plot shows the power spectra ($\int
P(\phi,f) d\phi $) and the second plot shows the pulsar amplitude
($\int P(\phi,f)df$).  The effect of the RFI in the unfiltered is
evident in increased noise throughout the pulsar phase.  The third
plot shows the pulse peak amplitude as a function of frequency to show
the RFI effect in more detail ($\int^{0.55}_{0.45} P(\phi,f) d\phi $).
The full impact of the RFI is shown in the fourth plot, the
signal-to-noise ratio in the peak amplitude.

Outside the RFI range the filtered and unfiltered data are identical
(to within the re-sampling error).  In the RFI range the filtered data
are consistent with interpolations from the adjacent channels. For these
observations the reference signal was obtained from the 3.5m antenna.

\subsection {Additional RFI Mitigation Problems}

At this stage in our investigations we conclude that the adaptive
filter could provide a solution to the RFI mitigation question
provided that we can master the apparently random and uncorrelated
power variations in the RFI that are shown in figure~\ref{filt2}.

We suggest that the underlying mechanism is multipath propagation
 which leads to
destructive interference.  We are exploring a number of counter-measures:

\begin{itemize}
\item Installing a larger antenna which could raise the signal power and reduce the
number of multi-path rays.

\item Installing a second reference antenna displaced from the first to provide some
spatial diversity.  We would need to add an intelligent IF switch to
select the better of the two filtered IFs.

\item Exploiting the known coding algorithm of the digital TV to reconstruct
a better reference signal with higher (and stable) INR.  A variant of
this strategy has been demonstrated as way to mitigate GLONASS RFI by
Ellingson et al. (2001).
\end{itemize}

\subsection{Sampling Issues}

Two-bit sampling has not been a limitation to these experiments.  Even
though the RFI is substantially larger than the receiver noise, it is
confined to about 10\% of the receiver bandpass and does not dominate
the noise presented to the sampler.  We suffer a further penalty in
the resampling stage at the filter output. However, the present
experiment suggests that such loss is modest. Care is also required in
the resampling stage since the filtered data may retain echoes of the
original two-bit sampling; a poor setting of the thresholds can negate
much of the filter's action.

As pulsar observers lock the sampler thresholds during an observation,
non-stationarity might seem to be an additional potential source of
trouble.  However, the adaptive filter should be neutral to this
provided that the relation between measured and true correlation
coefficients remains linear.  See Jenet and Andersen (1998) for a
detailed discussion of this issue.

\section{Conclusion}

The results of this program are encouraging - we believe that the
adaptive filter will be able to provide a satisfactory level of RFI
mitigation.  We will shortly move to a hardware implementation of the
filter.

It is clear that the mitigation can only succeed if we have a high
quality copy of the interference; we will need to devote more time and
effort into this side of the problem in the coming months.

\begin{acknowledgements}

We are grateful to Dr. S. Ord (Swinburne University) and
Dr. J. Reynolds (ATNF) for valuable advice and assistance in the
course of this program.  We acknowledge the support of the observatory
staff at Parkes.  The Parkes radio telescope is part of the Australia
Telescope which is funded by the Commonwealth of Australia for
operation as a National Facility managed by CSIRO.

\end{acknowledgements}

\newpage

\end{article}

\end{document}